\documentclass[12pt]{elsarticle}
\makeatletter
\def\ps@pprintTitle{%
 \let\@oddhead\@empty
 \let\@evenhead\@empty
 \def\@oddfoot{}%
 \let\@evenfoot\@oddfoot}
\makeatother
\usepackage{amssymb,amsmath,bm,mathtools,xcolor,float}
\numberwithin{equation}{section}
\usepackage[ruled,linesnumbered,vlined]{algorithm2e}
\usepackage{graphicx}
\usepackage[colorlinks=true,linkcolor=blue]{hyperref}
\usepackage{geometry}
\usepackage{subcaption}	
\usepackage{lineno}
\usepackage{siunitx}
\begin{document}

\captionsetup[subfigure]{skip=1pt, singlelinecheck=false}

\title{Shack-Hartmann wavefront sensing: A new approach to time-resolved measurement of stress intensity during dynamic fracture of small brittle specimens}

\author[1]{Liuchi Li\corref{cor1}}
\ead{lli128@jhu.edu}
\cortext[cor1]{Corresponding author}
\author[1,2]{Velat Kilic}
\author[1,2]{Milad Alemohammad}
\author[1,3,4]{K.T. Ramesh}
\author[1,2]{Mark A. Foster}
\author[1,3,4]{Todd C. Hufnagel\corref{cor1}}
\ead{hufnagel@jhu.edu}

\address[1]{Hopkins Extreme Materials Institute, Johns Hopkins University, Baltimore, MD 21218, USA}
\address[2]{Department of Electrical and Computer Engineering, Johns Hopkins University, Baltimore, MD 21218, USA}
\address[3]{Department of Materials Science and Engineering, Johns Hopkins University, Baltimore, MD 21218, USA }
\address[4]{Department of Mechanical Engineering, Johns Hopkins University, Baltimore, MD 21218, USA}

\begin{abstract}
The stress intensity factor describes the stress state around a crack tip in a solid material and is important for understanding crack initiation and propagation. Because stresses cannot be measured directly, the characterization of the stress intensity factor relies on the measurement of deformation around a crack tip. Such measurements are challenging for dynamic fracture of brittle materials where the deformation is small and the crack tip velocity can be high ($>\SI{1}{km.s^{-1}}$). Digital gradient sensing (DGS) is capable of full-field measurement of surface deformation with sub-microsecond temporal resolution, but it is limited to centimeter-scale specimens and has a spatial resolution of only $\sim\SI{1}{mm}$. This limits its ability to measure deformations close to the crack tip. Here, we demonstrate the potential of Shack-Hartmann wavefront sensing (SHWFS), as an alternative to DGS, for measuring surface deformation during dynamic brittle fracture of millimeter-scale specimens. Using an opaque commercial glass ceramic as an example material, we demonstrate the capability of SHWFS to measure the surface slope evolution induced by a propagating crack on millimeter-scale specimens with a micrometer-scale spatial resolution and a sub-microsecond temporal resolution. The SHWFS apparatus has the additional advantage of being physically more compact than a typical DGS apparatus. We verify our SHWFS measurements by comparing them with analytical predictions and phase-field simulations of the surface slope around a crack tip. Then, fitting the surface slope measurements to the asymptotic crack-tip field solution from linear elastic fracture mechanics, we extract the evolution of the apparent stress intensity factor associated with the propagating crack tip. We conclude by discussing potential future enhancements of this technique and how its compactness could enable the integration of SHWFS with other characterization techniques including x-ray phase-contrast imaging (XPCI) for multi-modal characterization of dynamic fracture.

\end{abstract}

 \begin{keyword}
Dynamic fracture, Brittle material, Shack-Hartmann wavefront sensor, Surface slope measurement, Stress intensity factor, Phase-field simulation
 \end{keyword}
 
 \maketitle
%\linenumbers

\section{Introduction}
The stress intensity factor (SIF or $K_\mathrm{I}$ in Mode-I condition) is a critical parameter for understanding the fracture of materials. Although it cannot be measured directly, the stress intensity factor can be obtained indirectly from measurements of displacements around a crack tip. An ideal measurement technique would provide full-field displacement measurement with good spatial resolution, be fast enough to allow measurement of the stress intensity around the tip of a propagating crack, and be suitable for use on an arbitrary material. Particularly challenging are studies of brittle materials, where the in-plane displacements around the crack tip can be small (on the order of tens of nanometers) and the crack-tip speed is high ($>\SI{1}{km.s^{-1}}$).

Over the past fifty years, extensive research has resulted in the development of a series of increasingly powerful optical techniques for characterizing the stress intensity of propagating cracks in a variety of materials. Notable examples include measurements relying on the photoelastic effect~\cite{dally1979dynamic}, the method of caustics~\cite{ravi1982experimental, chandar1982dynamic, rosakis1984determination}, coherent gradient sensing (CGS)~\cite{tippur1991coherent,tippur1991optical, krishnaswamy1992measurement,butcher1998functionally}), digital image correlation (DIC)~\cite{kirugulige2007measurement,albertini2021effective}), and, most recently, digital gradient sensing (DGS)~\cite{periasamy2012full, miao2018higher,sundaram2018dynamic}. Any of these techniques may be effective, depending on the material under examination, the specimen geometry, and the loading conditions, but all have limitations~\cite{tippur1991coherent, periasamy2012full}.

Measuring the stress intensity during dynamic fracture of brittle materials (defined here as materials showing little or no plastic deformation under the loading conditions of interest) is especially challenging because the displacements around the crack tip are small, the time available for the measurement is limited by the high crack tip speeds (often in excess of \SI{1}{km.s^{-1}}), and there is the potential for crack branching~\cite{dondeti2020comparative}. One of the most powerful techniques is digital gradient sensing (DGS), which can provide a full-field measurement around a dynamically propagating crack tip for both transparent and opaque brittle materials~\cite{periasamy2012full,dondeti2020comparative}. DGS has been applied to the study of dynamic fracture of a variety of materials including polymers~\cite{ sundaram2017dynamic},  oxide glasses~\cite{dondeti2022cascading}, and polymer-matrix composites~\cite{hao2015experimental, miao2019fracture}. 

When performed in reflection from an opaque material (or transparent material with a reflective coating), DGS measures the deflection of light rays due to the surface profile and, in particular, the slope associated with out-of-plane displacements due to stress concentration around the crack tip. From these displacements, the stress intensity factor can be extracted using fracture mechanics theory and the instant crack velocity~\cite{periasamy2013full}. DGS has an advantage over DIC for measurements on brittle materials, in that a long optical path (several meters) amplifies the small displacements enabling higher sensitivity. The field of view of DGS measurements is typically on the order of a few centimeters, with a spatial resolution of around a millimeter~\cite{dondeti2020comparative, dondeti2022cascading,miao2020simplified}. These characteristics make it challenging to use DGS on small samples, and require a significant amount of laboratory space for implementation.

Here we propose a new approach (as an alternative to DGS) to measure surface profile gradients during dynamic fracture of brittle materials, based on Shack-Hartman wavefront sensing (SHWFS)~\cite{shack1971production, platt2001history}. Our technique retains the advantages of gradient-based methods but provides significantly improved spatial resolution ($\sim\SI{10}{\micro.m}$) while making the experimental setup more compact (with a total optical path of $<\SI{1}{m}$). This combination makes SHWFS ideal for multi-modal studies, particularly in combination with \mbox{x-ray} phase contrast imaging (XPCI) performed at synchrotron facilities~\cite{luo2012gas,parab2014observation,leong2018quantitative}. In this paper, we describe the SHWFS technique, apparatus, and data analysis; illustrate its application to the study of dynamic fracture of a commercial glass ceramic (including extraction of the apparent stress intensity factor); and discuss potential extensions and enhancements of the technique.

\section{Shack-Hartmann wavefront sensor (SHWFS)}
\label{section:shwfs} % there is a figure named shwfs as well which causes issues

\begin{figure}
\centering
\includegraphics[width=\linewidth]{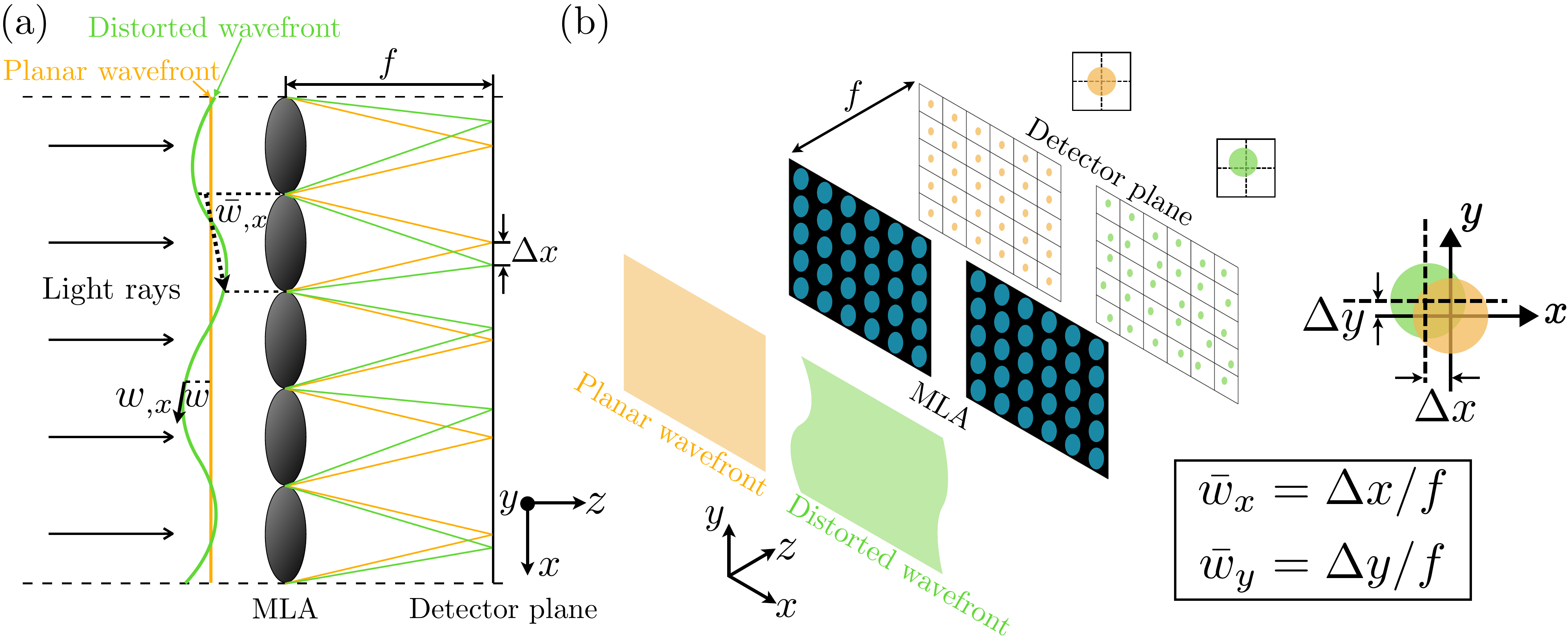}
\caption{(a) A 2D schematic along the $x$ direction demonstrating the working principle of SHWFS. The distorted wavefront goes through a collection of microlenses (with a uniform focal length $f$), each of which performs a lens-area-averaged estimation of the wavefront's gradient, denoted as $\bar{w}_{x}$. (b) A 3D schematic demonstrating the working principle of SHWFS. }
\label{shwfs}
\end{figure}

\subsection{Working principle}
The Shack-Hartman wavefront sensor (SHWFS) was originally developed to measure distortions in telescope images due to atmospheric turbulence and enabled the field of adaptive optics~\cite{platt2001history}. Here we use SHWFS in a manner similar to reflection-mode DGS to measure surface slopes and extract fracture parameters during dynamic fracture. The major difference from DGS is in how the surface slope is measured. In DGS, an image of a reference pattern is recorded in reflection off of the sample; when the sample is deformed, the pattern becomes distorted relative to the original image, and from the distortion, the surface gradients are determined. In SHWFS, instead of imaging a reference pattern, a microlens array (MLA) is used to measure distortions of a planar wavefront reflected from the sample.

A basic SHWFS consists of a micro-lens array and an image sensor, as shown in Fig.~\ref{shwfs}. With the sensor positioned at the focal plane of the MLA and illuminated by a plane wave, each micro-lens creates a focused spot on its respective optical axis. A distorted wavefront on the other hand produces a focused spot displaced from the optical axis. The displacement of the centroid of each spot is then proportional to the orientation of the wavefront averaged across the aperture of the corresponding microlens. The microlenses thus map angular deformations of the wavefront to centroid shifts at the focal plane as depicted in Fig.~\ref{shwfs}(a):
\begin{align}\label{wx}
\bar{w}_{x} \approx \Delta x/f,
\end{align}
where $\bar{w}_{x}$ is the averaged gradient of the wavefront across a particular microlens, $\Delta x$ is the displacement of the centroid on the detector plane, and $f$ is the focal length of the microlens. Fig.~\ref{shwfs}(b) extends Fig.~\ref{shwfs}(a) to a 3D schematic, where we can also compute similarly the gradient in the transverse direction, 
\begin{align}\label{wy}
\bar{w}_{y} \approx \Delta y/f.
\end{align}
In essence, by sampling the wavefront with an MLA all of these shifts can be measured, and the whole wavefront can be reconstructed provided the scene is properly sampled both spatially and temporally \cite{schmutz1987hartmann}, see section. \ref{shwfs_fracture} for a more detailed discussion.% The requirements are as follows:

\subsection{Application in the context of dynamic fracture}\label{shwfs_fracture}
We use one surface of a sample to generate a wavefront by reflecting off incoming collimated light rays. For this work, we consider only the Mode-I loading condition and use the single-notched three-point bend configuration for our experiment. As shown in Fig. \ref{shwfs_sample}(a), initially, when the sample is stress-free, the surface is perfectly flat, giving a planar wavefront shown as a set of equally spaced spots on the detector plane. Next, when the sample is loaded in-plane, due to Poisson's effect and heterogeneous in-plane stress distribution, the sample surface will deform out-of-plane heterogeneously, leading to non-zero surface slope and shifts of those spots on the detector plane, as shown in Fig. \ref{shwfs_sample}(b). During a dynamic fracture event, a propagating crack at every time instant will cause a spatial variation of the surface slope (especially near the crack tip due to stress concentration), which we calculate by first measuring the shifts $\Delta x$ and $\Delta y$ and next feeding into Eqns. \ref{wx} and \ref{wy}. For successful detection of $\Delta x$ and $\Delta y$, the following requirements should be met:

\begin{figure}
\centering
\includegraphics[width=\linewidth]{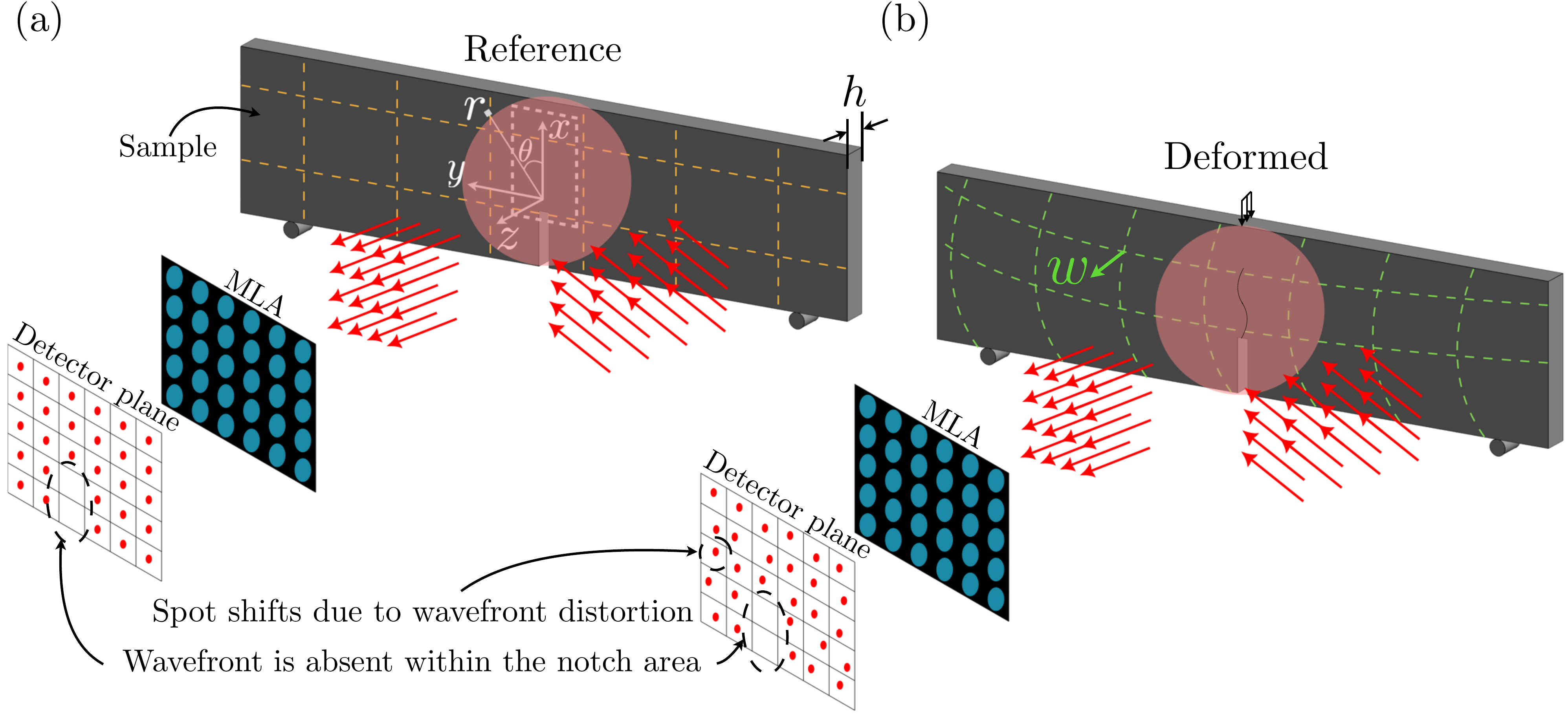}
\caption{Application of SHWFS in the context of dynamic fracture using a single-notch three-point bending configuration. The planar (reference) wavefront is generated by the sample's smooth surface reflecting incoming collimated light rays, shown in (a). In contrast, the distorted (deformed) wavefront is caused by the surface deforming out-of-plane due to Poisson's effect, as shown in (b). In (a), the polar coordinate associated with the crack tip is also shown.}
\label{shwfs_sample}
\end{figure}  

\begin{itemize}
    \item \emph{Spatial sampling} --- The MLA pitch (the distance from the center of one lenslet to the other) should be smaller than the spatial correlation length of the distorted wavefront so that the wavefront can be safely assumed to be locally planar. Indeed, any function can be approximated locally linearly, provided it is sampled sufficiently densely. Although a smaller pitch length is preferred, we note that diffraction puts a lower bound on the pitch length for an MLA to function properly. Thus, we can think of the pitch length as similar to the correlation window size in DIC: a large window size reduces measurement noise but is prone to average out locally high deformation gradients, the opposite is true for a small window size. There is also a lower bound for the correlation window size with regard to accurate measurements, which is related to the average speckle feature size and contrast \cite{pan2008study}.

    \item \emph{Temporal sampling} --- In dynamic fracture experiments the sample surface can deform significantly and rapidly. To measure surface distortions at a given time instant, the incident wavefront must be essentially constant, which requires that it be sampled at a rate faster than its correlation time. If the incident wavefront varies significantly over a single exposure time of the recording camera, the focused spot will move as a result, and the integrated image will be blurred. The same principle also applies to DIC analysis (and any other imaging techniques), where the in-plane deformation can be viewed as the incident wavefront, whose variation should not be significant within a single exposure time of the recording camera. Therefore, a high-speed camera is needed. For imaging dynamic fracture in glass and ceramics, a frame rate on the order of one million per second is sufficient \cite{dondeti2020comparative, dondeti2022cascading}.

    \item \emph{Spot size and shape} ---
    A tightly focused spot array on the detector plane is essential for accurate detection, and it ideally requires collimated illumination from a spatially confined source (fiber-coupled laser in our case), a smooth sample surface, and a large enough lenslet diameter. A collimated source is necessary to achieve a tightly focused spot array since the source plane is mapped onto the detector plane. However perfect collimation is not a strict requirement since the spot positions are measured differentially (i.e., referenced with respect to the positions before crack initiation). A smooth sample surface ensures that the in-coming planar (i.e., collimated) illumination remains largely unchanged after reflection. In practice, a mirror-like finish (with a roughness average $R_a < \SI{0.3}{\micro m}$) is sufficient for our purpose. Lastly, a small lenslet will lead to a large (as opposed to tightly focused) spot shape in the Fourier plane due to diffraction, which is undesired.
    
    \item \emph{Dynamic range} --- The MLA pitch and focal length place a limit on the surface tilt that can be measured. In particular, the shift of the focused spots must be large enough to be measured, but not so large that spots from adjacent lenses impinge on one another. In that case, it becomes challenging to disambiguate whether a change in the centroid position is caused by a small deformation, or a deformation so large that the centroid is coupled to the neighboring lenslet area. This effectively puts a limit on the range of spot shifts that can be measured, with a lower bound set by the camera pixel size and an upper bound set by the MLA pitch. We note, though, that computational methods could be used to extend this range. Indeed, the spot centroids may be located with a sub-pixel resolution if the spot is spatially sampled sufficiently densely by the camera pixels and with a sufficient signal-to-noise ratio (SNR)~\cite{mortensen2010optimized}. 
\end{itemize}

\section{Sample preparation and experimental setup}\label{ep}
\subsection{Sample preparation}
We use Macor\textsuperscript{\texttrademark}, a commercially available glass ceramic, as an example material to test the utility of SHWFS because its mechanical properties (such as the fracture toughness $K_\text{IC}$ and Young's modulus $E$) are well characterized and because it is highly machinable, making sample fabrication convenient. We purchased Macor\textsuperscript{\texttrademark}\textsuperscript{\texttrademark} plates (Master-Carr 8489K231) with an initial thickness of \SI{0.0625}{in} (\SI{1.58}{mm}). We used a diamond wire saw machine to cut each plate into rectangular bars \SI{12}{mm} long (along the $y$ direction), \SI{3}{mm} wide (along the $x$ direction), and \SI{1.58}{mm} thick (denoted as $h$ in Fig.~\ref{shwfs_sample}(a), along the $z$ direction). We next used the same saw to create a  \SI{1}{mm}  deep notch in each bar to produce the single-notched three-point bending configuration mentioned in the previous section. The wire used has a diameter of \SI{250}{\micro m}, creating a notch width of \SI{250}{\micro m} and a semicircular notch tip. We next polished one face (the $x$-$y$ plane) of each bar to a mirror-like finish with diamond lapping films down to a \SI{0.2}{\micro m} grade). This forms the reflective surface necessary for the SHWFS approach. After polishing, specimens are found to have a reduced thickness of $1.5 \pm 0.2$ mm. Note that it is not necessary to deposit a reflective coating, at least not for these specimens. For transparent materials such as glass, a reflective coating is desired.

\subsection{Experimental setup}
We use a custom-designed loading apparatus~\cite{leong2018quantitative} to induce dynamic fracture under three-point bending. In this apparatus, an indenter is connected to a piezoelectric actuator (Cedrat Technologies PPA40M), with the actuator being driven by a voltage signal that is generated by a function generator (Tektronix AFG3252) and then amplified by a high-speed voltage amplifier (PiezoDrive PD200). We place each sample over a loading plate with rolling support that sits above a vertical translation stage (OptoSigma TSD-653DMUU). We use the function generator to output a single linear voltage ramp from \SI{0}{V} to \SI{7}{V} over a time window of \SI{256}{\micro s}, resulting in an indentation speed of approximately \SI{0.13}{m.s^{-1}}. (Details of the speed measurement are provided in the appendix.) This apparatus is compact enough for operations in synchrotron facilities, and it has been used to characterize dynamic fracture of different brittle materials via XPCI~\cite{leong2018quantitative, kang2020crack}. With this loading apparatus inducing crack propagation within a sample, we use a high-speed camera (Shimadzu HPV-X2) to capture the distorted wavefront (subsequently $\bar{w}_{x}$ and $\bar{w}_{y}$) evolution around the propagating crack tip.

\begin{figure}
\centering
\includegraphics[width=\linewidth]{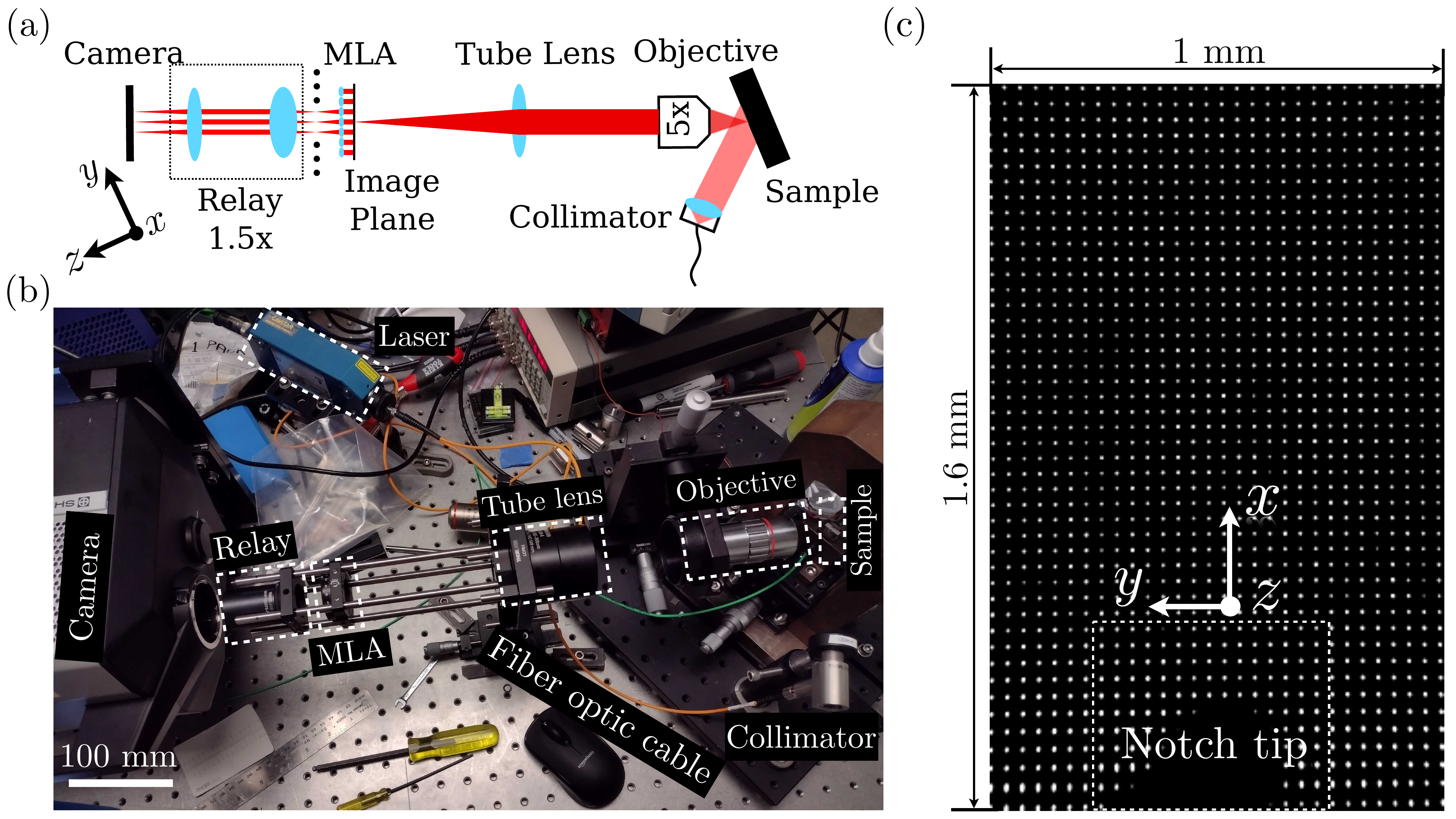}
\caption{(a)~Schematic of the SHWFS design. (b)~Top-down photograph of the SHWFS setup. (c)~Image formed on the camera detector showing the planar wavefront generated from the polished surface of the undeformed sample.}
\label{shwfs_setup}
\end{figure}

Figure~\ref{shwfs_setup} shows a schematic and a photograph of the optical setup. Illumination of the sample was provided by a pulsed laser beam (SI-LUX 640) with a fiber-coupled collimator (M92L01 and RC08FC-F01 from Thorlabs). The reflection from the sample was imaged with an infinity-corrected microscopy system (Mitutoyo $5\times$ objective with a ThorLabs TTL200-A tube lens), a microlens array (ThorLabs MLA150-5C-M with a pitch of $\SI{150} {\micro m}$), and a telescopic relay lens pair (ThorLabs MP105075-A) with $1.5\times$ magnification. Note the compactness of the experimental setup, which fits in considerably less space than typically required for DGS. A typical spot pattern from an undeformed specimen is shown in Fig.~\ref{shwfs_setup}, where the field of view is approximately $\SI{1.6}{mm}\times \SI{1}{mm}$, with the spacing between spots about \SI{45}{\micro m}. The spot spacing can be calculated using the pitch of the microlens array and the magnification ratio of the microscope and the relay: $\SI{150}{\micro m}/5\times 1.5=\SI{45}{\micro m}$.

\section{Calculation of shift and surface slope from SHWFS recording} \label{calculation}
\subsection{Shift and surface slope}
\begin{figure}
\centering
\includegraphics[width=\linewidth]{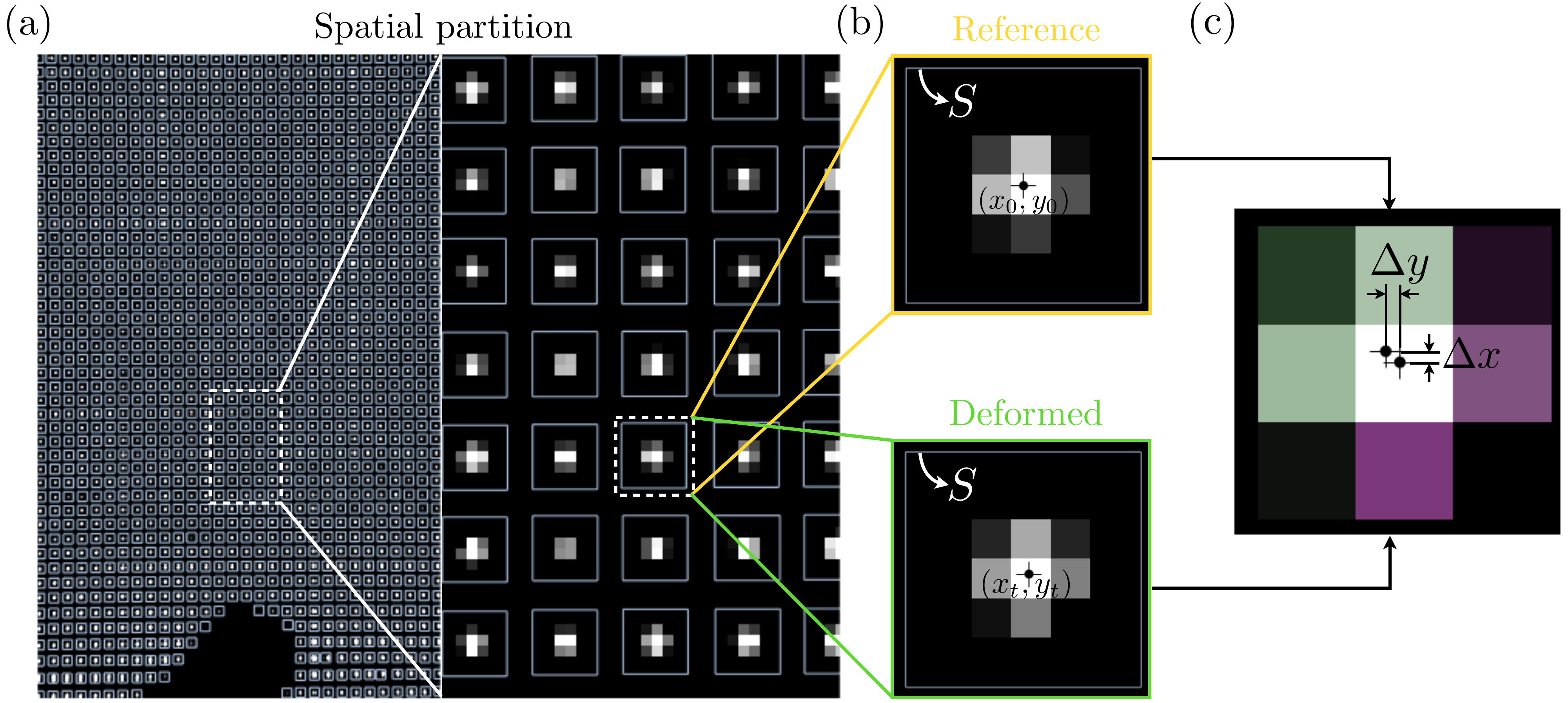}
\caption{(a)~Spatial partitioning of light spots, each assigned to a unique square cell. (b)~Examples of calculating the centroid of one spot in the reference (top) and deformed (bottom) configurations. (c)~An example of calculating the shifts $\delta x$ and $\delta y$ by overlaying the two images from (b). Areas colored in either green or purple are mismatches between the two images.}
\label{shwfs_calculation}
\end{figure}

As a crack initiates and propagates through a sample, it will cause spots focused on the image plane to shift laterally. (Refer to the Supplemental Material for a representative recording.) We measure these shifts and later use them to extract the apparent stress intensity factor (SIF). For a particular recording, we start from the initial reference image and perform a spatial partition such that each focused spot belongs to one and only one cell, as shown in Fig.~\ref{shwfs_calculation}(a). For images captured after  crack initiation, we use this partition to calculate the shift $\Delta x$ and $\Delta y$ of each spot by finding its centroid in the reference (denoting as $[x_0, y_0]$) and the deformed configuration at a particular time instance $t$ (denoting as $[x_t, y_t]$):
\begin{align}\label{delta}
[\Delta x, \Delta y]|_t = [x_t-x_0, y_t-y_0].
\end{align}

The procedure for finding the centroid of each spot is illustrated in Fig.~\ref{shwfs_calculation}. For a given spot we calculate the intensity-weighted average of all pixels that constitute that spot, in a way similar to Ref. \cite{baik2007center}. Taking one spot in the reference image as an example, its centroid is given by
\begin{align}\label{centroid}
x_0 = \frac{\sum_{i,j \in S }I_{ij}\times i}{\sum_{i,j\in S}I_{ij}}, y_0 =  \frac{\sum_{i,j \in S }I_{ij}\times j}{\sum_{i,j\in S}I_{ij}},
\end{align}
where $(i, j)$ denotes the location of a pixel, $S$ denotes the cell captured by an individual lenslet, and $I_{ij}$ denotes the intensity of the pixel at location $(i, j)$. We then apply Eqn.~\ref{centroid} to find $(x_t, y_t)$ for the same spot in each subsequent image (Fig.~\ref{shwfs_calculation}(b)). Lastly, we calculate $\Delta x$ and $\Delta y$ using Eqn.~\ref{delta}, as shown in Fig.~\ref{shwfs_calculation}(c). 

The next step in the data reduction is converting the spot shifts ($\Delta x$ and $\Delta y$) to surface slopes ($\bar{w}_{x}$ and $\bar{w}_{y})$. Care must be taken in this process because the microscope provides transverse magnification for increased spatial resolution, but this leads to angular de-magnification which in turn reduces the sensitivity to out-of-plane deformations. In Fourier optics terms, magnifying a 2D signal must necessarily reduce its spatial bandwidth. On the other hand, the relay used after the MLA images the focal plane of the MLA and increases system sensitivity to out-of-plane deformations. Putting these effects together,
\begin{align}\label{conversion}
\frac{s_r}{s_m}\bar{w}_{x} = \frac{d\,\Delta x}{f}\qquad \text{and}\qquad \frac{s_r}{s_m}\bar{w}_{y} = \frac{d\,\Delta y}{f},
\end{align}
where $s_m$ and $s_r$ are the magnification factors of the microscope objective ($s_m= 5$ here) and the relay lens pair ($s_r= 1.5$ here), $d$ is the pixel size of the camera sensor ($d= 30\,\, \mu$m), $\Delta x$ (or $\Delta y$) is the shift  in pixels, and $f$ is the focal length of the MLA ($f= \SI{5.6}{mm}$ here). In contrast, if we want to calculate in-plane information such as the distance $L$ between any two pixels $(i_1, j_1)$ and $(i_2, j_2)$, we will have:
\begin{align}\label{inplane}
s_rs_mL= d \sqrt{(i_1-i_2)^2+(j_1-j_2)^2}.
\end{align}

\subsection{Extraction procedure for $K_\text{I}$ using slope data}\label{extraction_prodecure}
Once we have measured the shifts $\Delta x$ and $\Delta y$ of each spot and computed the corresponding surface slope $\bar{w}_{x}$ and $\bar{w}_{y}$ using Eqn.~\ref{conversion}, we can extract the apparent stress intensity factor $K_\text{I}$ by fitting the slope data to the asymptotic crack-tip stress field \cite{williams1957stress,freund1998dynamic} for a Mode-I crack opening under plane stress condition:
\begin{align}\label{slopeK_full}
\begin{split}
 -\frac{2E}{\nu h}\bar{w}_{x} = \frac{\partial \left( \hat{\sigma}_{xx}+\hat{\sigma}_{yy}\right)}{\partial x} =-\frac{1}{2}r^{-\frac{3}{2}}g(V)A_1\text{cos}\left( \frac{3}{2}\theta \right)+\sum_{N=2}^{\infty}A_N\left(\frac{N}{2}-1 \right)r^{\left(\frac{N}{2}-2\right)}\text{cos}\left[\left(\frac{N}{2}-2 \right) \theta\right],\\ 
 -\frac{2E}{\nu h}\bar{w}_{y} = \frac{\partial \left( \hat{\sigma}_{xx}+\hat{\sigma}_{yy}\right)}{\partial y} =-\frac{1}{2}r^{-\frac{3}{2}}g(V)A_1\text{sin}\left( \frac{3}{2}\theta \right)+\sum_{N=2}^{\infty}A_N\left(\frac{N}{2}-1 \right)r^{\left(\frac{N}{2}-2\right)}\text{sin}\left[\left(\frac{N}{2}-2 \right) \theta\right],
 \end{split}
\end{align}
where $\nu, E, h$ are the Poisson's ratio, the Young's modulus, and the thickness of the sample, respectively; $\hat{\sigma}_{xx}$ and $\hat{\sigma}_{yy}$ are the thickness averages of stress components along the $x$ and $y$ directions; $(r,\theta)$ denotes the position relative to the crack-tip in polar coordinates (see Fig. \ref{shwfs_sample}(a)), where $r$ at each data point can be computed following Eqn. \ref{inplane}; $A_1 = K_\text{I}\sqrt{\frac{2}{\pi}}$ where $K_\text{I}$ is the Mode-I stress intensity factor (SIF); $V$ is the crack tip speed with $g(V)$ being a function of the instantaneous crack tip speed to account for the velocity dependency of the K-field \cite{freund1998dynamic}. It takes the following expression:
\begin{align}
g(V) = \frac{1+\nu}{1-\nu} \frac{\left(1-\alpha_L^2\right)\left(1+\alpha_S^2\right)}{4\alpha_L\alpha_S-\left(1+\alpha_S^2\right)^2}, \,\, \text{with}\,\, \alpha_L^2 = 1-\frac{V^2}{C_L^2} \,\,\text{and}\,\,  \alpha_S^2 = 1-\frac{V^2}{C_S^2},
\end{align}
where $C_L$ and $C_S$ are longitudinal and shear wave speeds in the material. For relatively slow cracks ($V < 0.4C_R$ where $C_R$ is the Rayleigh wave speed  \cite{svetlizky2014classical}), $g(V) \simeq 1$ and the velocity dependence is negligible~\cite{freund1998dynamic}. Because our experiments are conducted within this slow crack velocity regime ($V \sim \SI{1e2}{m.s^{-1}}$ while $C_R>\SI{1e3}{m.s^{-1}}$), we take $g(V) = 1$ and neglect higher-order terms in Eqn.~\ref{slopeK_full}, resulting in the following expressions:
\begin{align}\label{slopeK_simple}
\begin{split}
-\frac{2E}{\nu h}\bar{w}_{x} = \frac{\partial \left( \hat{\sigma}_{xx}+\hat{\sigma}_{yy}\right)}{\partial x} \simeq-\frac{K_\text{I}}{\sqrt{2\pi}}r^{-\frac{3}{2}}\text{cos}\left( \frac{3}{2}\theta \right),\\
-\frac{2E}{\nu h}\bar{w}_{y} = \frac{\partial \left( \hat{\sigma}_{xx}+\hat{\sigma}_{yy}\right)}{\partial y} \simeq-\frac{K_\text{I}}{\sqrt{2\pi}}r^{-\frac{3}{2}}\text{sin}\left( \frac{3}{2}\theta \right).
\end{split}
\end{align}

\begin{figure}
\centering
\includegraphics[width=\linewidth]{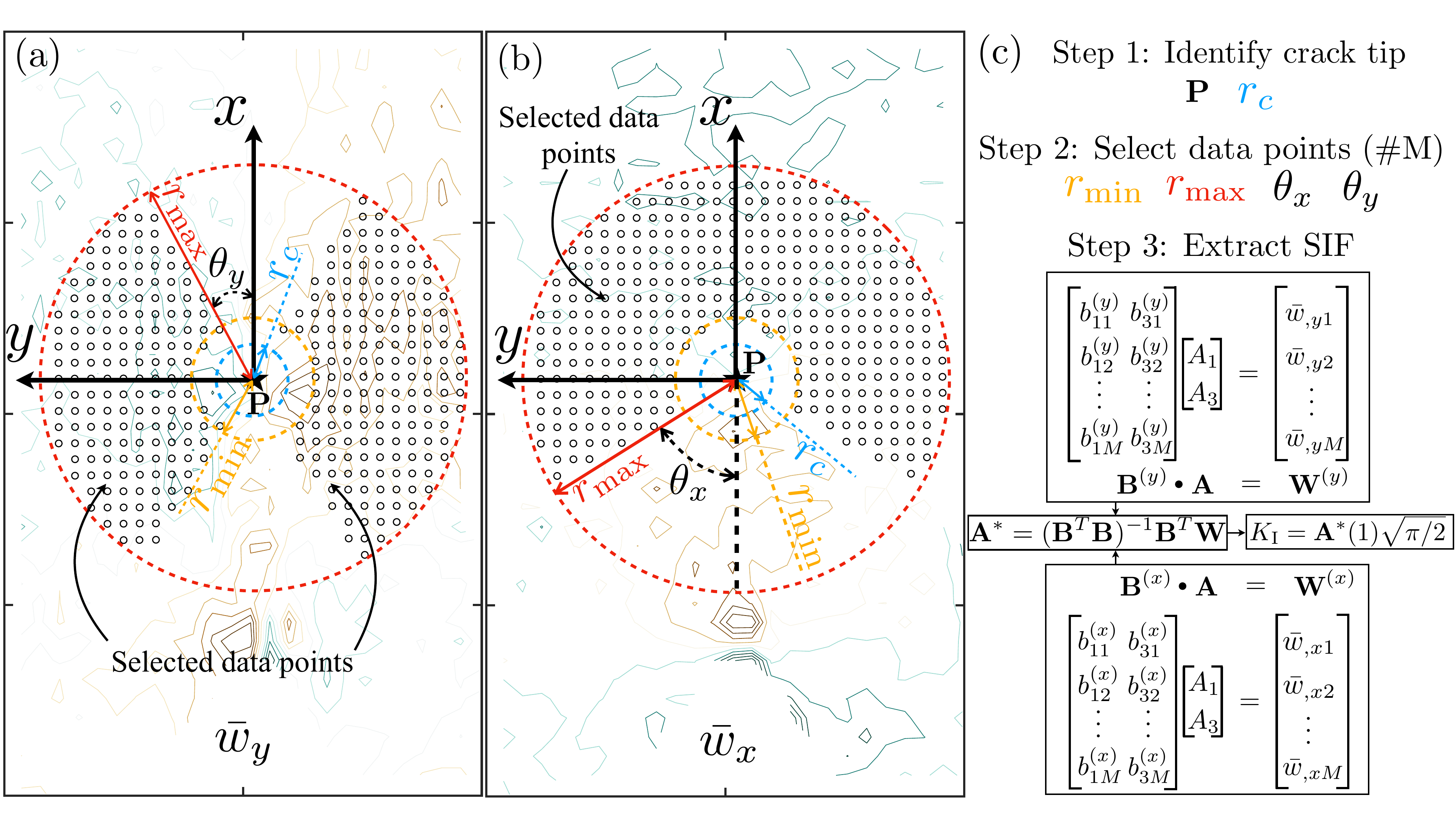}
\caption{Selection of points for determining $K_\mathrm{I}$ from surface slope data. Parts (a) and (b) show contour plots of $\bar{w}_{y}$ and $\bar{w}_{x}$, respectively (see~\ref{result_slope} for a description of these data). The superimposed circles represent positions from which the surface slope data were used to determine $K_\mathrm{I}$. The region of data used is specified by four parameters: the position of the crack tip, $\mathbf{P}$; inner and outer radii, $r_\text{min}$ and $r_\text{max}$; and the half-angle of an excluded region,  $\theta_y$ and $\theta_x$. The parameter $r_c$ reflects uncertainty in determining $\mathbf{P}$, as described in the text. (c)~A flow chart showing steps for calculating the apparent stress intensity factor $K_\text{I}$.
\label{data_fit}}
\end{figure}

With these expressions in hand, we can determine the stress intensity ($K_\text{I}$) during crack propagation from the surface slope data using an iterative procedure.  Our approach to data selection and fitting is similar to that presented in Ref.~\cite{dondeti2022cascading}, and has three basic steps:
\begin{itemize}
\item Identify the crack tip location $\textbf{P}$ based on the spatial distribution of  $\bar{w}_{x}$ and $\bar{w}_{y}$. We use a parameter $r_c$ to assess the uncertainty associated with $\textbf{P}$, as shown in Figs. \ref{data_fit}(a) and (b);
\item Choose a subset of the $\bar{w}_{x}$ and $\bar{w}_{y}$ data to use in determining $K_\text{I}$. The choice is described by four parameters: $r_\text{min}$, $r_\text{max}$, $\theta_x$, and $\theta_y$. The points chosen for $\bar{w}_{y}$ (Fig.~\ref{data_fit}(a)) correspond to $r_\text{min} \leq r \leq r_\text{max}$ and either  $\theta_y \leq \theta \leq \pi-\theta_y$ or $\pi+\theta_y \leq \theta \leq 2\pi-\theta_y$. The points chosen for $\bar{w}_{x}$ (Fig.~\ref{data_fit}(b)) are similar but with a different range of angles (specified by $\theta_x$ rather than $\theta_y$). Note that the number of points chosen for $\bar{w}_y$ and $\bar{w}_x$ may be different in general. 
\item Fit the data from the selected points using Eqn.~\ref{slopeK_full} to determine $K_\text{I}$. To do so, we first calculate the coefficients (denoted as $b_1$ and $b_3$) associated with the first two nonzero terms in Eqn.~\ref{slopeK_full} (\emph{i.e.} those with $A_1$ with and $A_3$) for each data point selected from the previous step. $b_1$ and $b_3$ are determined from the geometrical location ($r$ and $\theta$) of each selected point. Once we have these coefficients, we perform a least-squares minimization to find the vector $\mathbf{A} = [A_1, A_3]^{T}$ (subsequently extracting $K_\text{I} = \sqrt{\pi/2}A_1$), whose solution is given by the pseudo inverse of $\mathbf{B}$ shown in Fig.~\ref{slopeK_full}(c): 
\begin{align}\label{minimization}
\mathbf{A}^*  = \min_{\mathbf{A}} \left \lVert  \mathbf{B}^{(y)}\mathbf{A}-\mathbf{W}^{(y)} \right \lVert \quad \text{or} \quad \mathbf{A}^*  = \min_{\mathbf{A}} \left \lVert  \mathbf{B}^{(x)}\mathbf{A}-\mathbf{W}^{(x)}  \right \lVert,
\end{align}
where $\mathbf{B}^{(y)}$ (or $\mathbf{B}^{(x)}$) is a matrix containing $b_1$ and $b_3$, and $\mathbf{W}^{(y)}$ (or $\mathbf{W}^{(x)}$) is a vector containing the surface slope data for the $y$ direction (or the $x$ direction), as shown in Fig.~\ref{data_fit}(c). Alternatively, we can merge the two minimization problems into one: $\mathbf{A}^* =\min_{\mathbf{A}} \left \lVert [\mathbf{B}^{(y)};\mathbf{B}^{(x)}]\mathbf{A}-[\mathbf{W}^{(y)};\mathbf{W}^{(x)}]\right \lVert$. This fitting procedure using the out-of-plane information is similar to that which uses in-plane information (such as from DIC measurements) to extract $K_\text{I}$ for polymers~\cite{albertini2021effective}. 
\end{itemize} 

As we can see from Eqn.~\ref{slopeK_full} the terms dominating the solution of Eqn.~\ref{minimization} are those from points close to the crack tip, which have large  $r^{-3/2}$ and $r^{-1/2}$ as well as large $\bar{w}_{x}$ and $\bar{w}_{y}$. This makes precise determination of the position of the crack tip ($\mathbf{P}$) critical. On the other hand, we want to exclude data that are too close to the crack tip and near the crack path, because the SHWFS measurements become unreliable there due to the highly localized deformation and discontinuity. Thus, selecting an appropriate value of  $r_\text{min}$ balancing these considerations is also important. The other parameters ($r_\text{max}$, $\theta_x$, and $\theta_y$) are much less important than $\mathbf{P}$ and $r_\text{min}$. 

In practice, we have found that choosing $r_\text{min} = 20$~pixels (\SI{0.08}{mm} on the specimen), which excludes data from the closest two or three spots to the crack tip, works well with our setup. Other parameters chosen are $r_\text{max}$= 110~pixels (equivalent to  \SI{0.4}{mm} on the sample, which covers essentially the whole horizontal field of view), $\theta_y = \pi/6$, and $\theta_x  = \pi/4$.  Lastly, we locate $\mathbf{P}$ for each snapshot based on the symmetric spatial pattern of $w_y$ and $w_x$ shown in Fig.~\ref{result_perp}(a) and Fig.~\ref{result_para}(a), in a way similar to Ref.~\cite{sundaram2018dynamic}. We denote this location of $\mathbf{P}$ as the base position, whose location is indicated by the black star in Figs.~\ref{data_fit}(a) and (b).

Of course, there is uncertainty associated with both $\mathbf{P}$ and $r_\text{min}$, especially the former, and it is important to quantify how this affects the value $K_\text{I}$ determined from the measurements. We assess the uncertainty in $\mathbf{P}$ by sampling multiple positions of $\mathbf{P}$ in a circle of a radius of $r_c$ centered at the base position of $\mathbf{P}$. Similarly, we apply a variation $r_d$ to the value of $r_\text{min}$. We calculate the value of $K_\text{I}$ over each of these ranges and report the average value. In practice, we have found that setting $r_c=r_d=5$~pixels (\SI{0.02}{mm}) works well. Finally, we do not analyze data from the initial ten images starting from crack initiation (i.e., crack lengths less than \SI{0.3}{mm}) due to the difficulty in identifying $\mathbf{P}$ for such short cracks. We note that this could be alleviated by using fatigue-precracked specimens, where the more pronounced stress concentration would aid in the determination of $\mathbf{P}$.

\section{Results}
\label{result} 
\subsection{Measurement of surface slope $\bar{w}_{x}$ and $\bar{w}_{y}$}\label{result_slope}
\begin{figure}
\centering
\includegraphics[width=\linewidth]{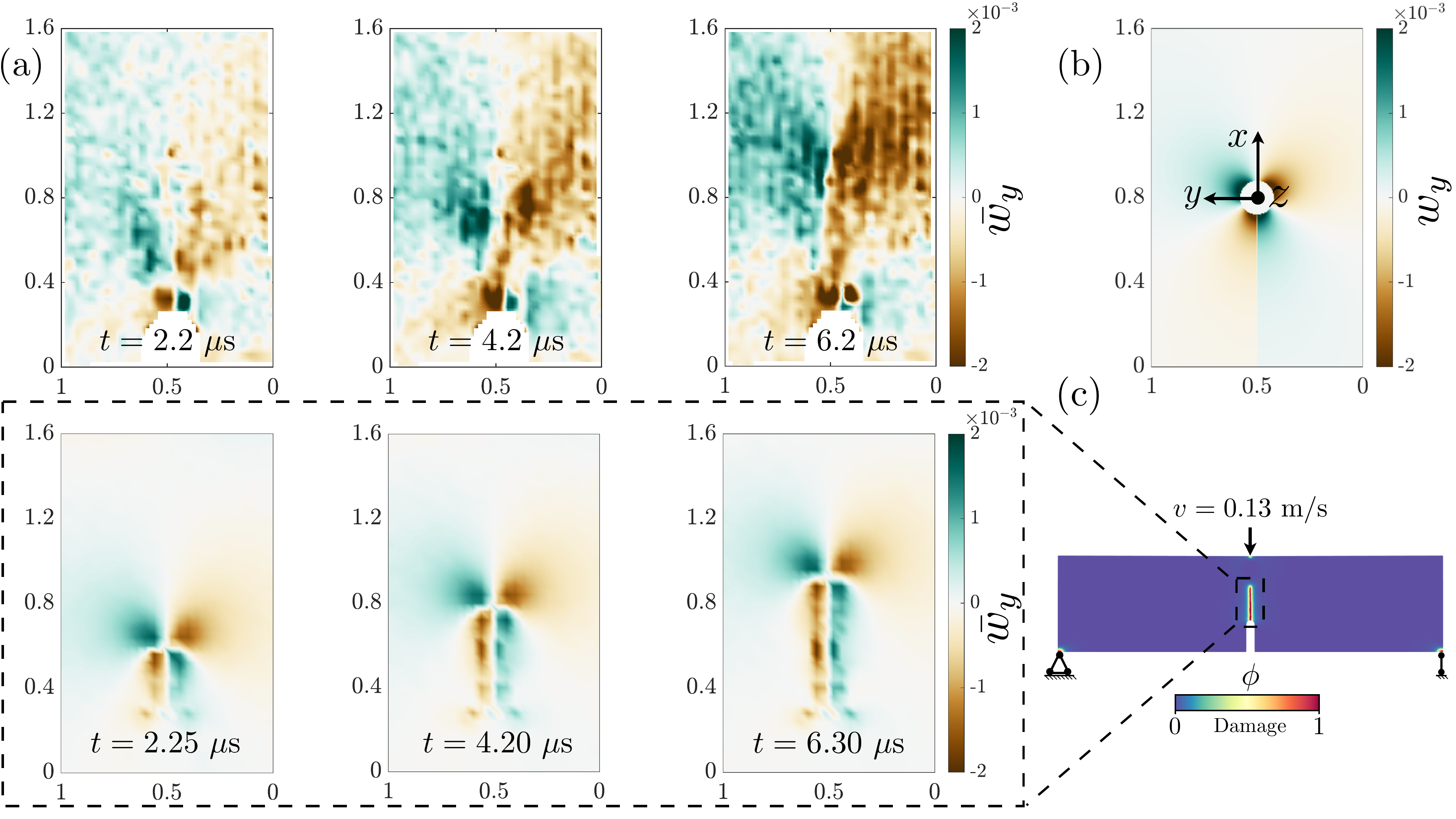}
\caption{(a) Spatial distribution of surface slope $\bar{w}_{y}$ determined from SHWFS measurements at three different times after crack initiation ($t=0$). (b)~Spatial distribution of $w_{y}$ calculated from the near-tip stress field (analytical solution), given a crack tip at the center of the image. (c)~A schematic showing the dynamic phase-field simulation and three plots showing the computed distribution of $\bar{w}_{y}$ at times comparable to the experimental data in (a). The unit for all color bars is radians, and that for all axes is millimeters.}
\label{result_perp}
\end{figure}

Fig.~\ref{result_perp}(a) shows the distribution of $\bar{w}_{y}$ across the field of view at three instants during a representative dynamic fracture experiment from a notched three-point bend sample of Macor\textsuperscript{\texttrademark}. We use the crack initiation time as a reference ($t = 0$). For this experiment with images collected at \SI{5}{Mfps}, the time between frames is \SI{0.2}{\micro s}. The spatial pattern (shape and magnitude) of $\bar{w}_{y}$ is consistent with a calculation (Fig.~\ref{result_perp}(b)) using Eqn.~\ref{slopeK_simple} and the mechanical properties of Macor\textsuperscript{\texttrademark} ($E = \SI{66.9}{GPa}$, $\nu = 0.29$, and $K_\text{IC} = \SI{1.53}{MPa.m^{1/2}}$). As an additional comparison we compute $\bar{w}_{y}$ (Fig.~\ref{result_perp}(c)) from dynamic phase-field  simulations~\cite{miehe2010phase, borden2012phase} of fracture using a custom code~\cite{li2023meso}. (Details of the phase-filed simulations are provided in the  Appendix). This computation is achieved by plugging in the in-plane stresses $\sigma_{xx}$ and $\sigma_{yy}$ obtained from the simulation into Eqn.~\ref{slopeK_simple}. We believe that the highly localized surface slope profile in the simulation results in the wake of the crack is an artifact of the phase-field algorithm, which smears the sharp crack over a certain region, in which the stress does not completely drop to zero.

\begin{figure}
\centering
\includegraphics[width=\linewidth]{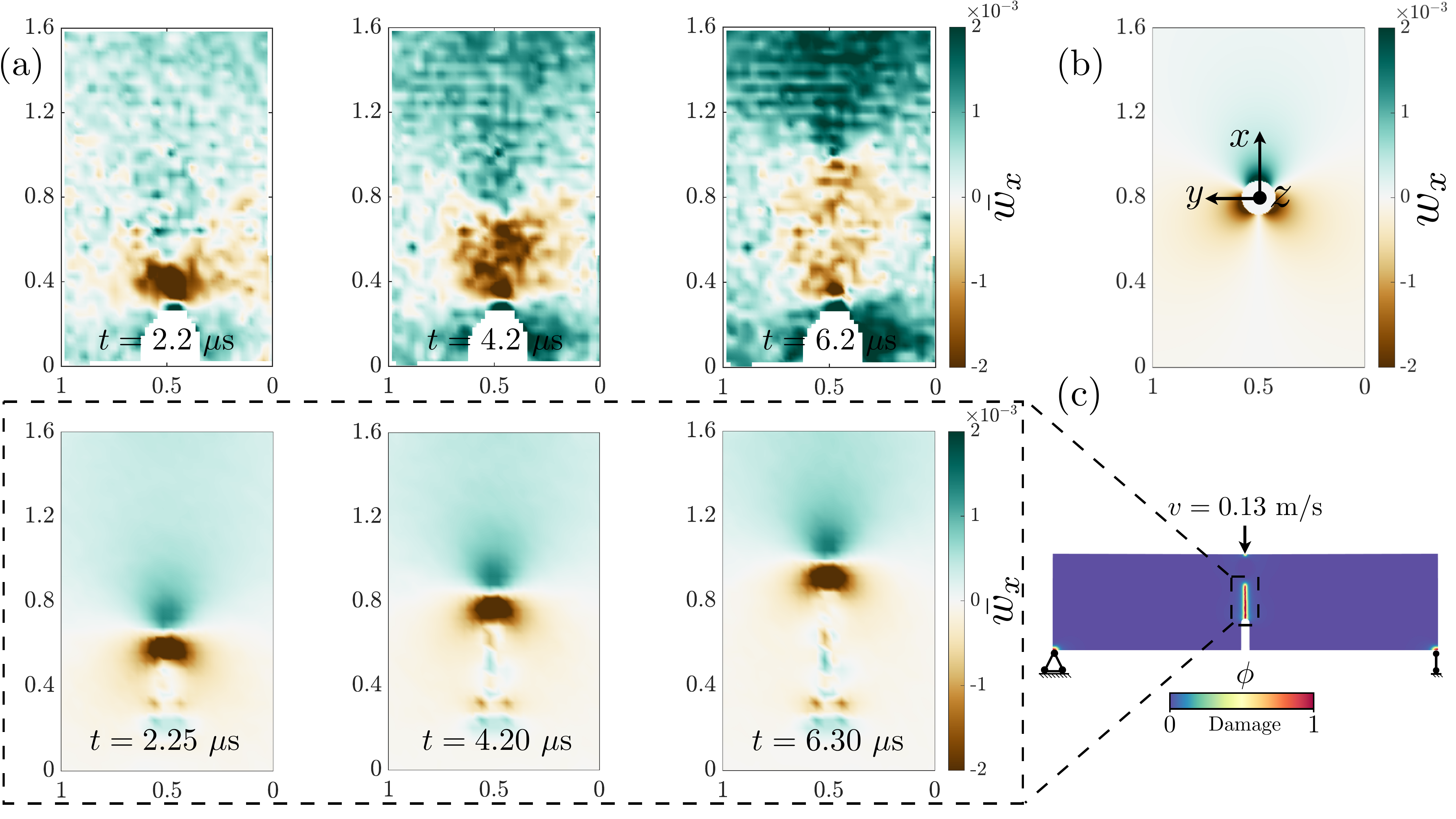}
\caption{Similar to Fig.~\ref{result_perp} but for surface slopes $w_{x}$. (a) Spatial distribution of $\bar{w}_{x}$ determined from SHWFS measurements. (b)~Analytical solution for spatial distribution of $w_{x}$. (c)~A schematic showing the dynamic phase-field simulation and three plots showing the computed distribution of $\bar{w}_{x}$. The unit for all color bars is radians, and that for all axes is millimeters.}
\label{result_para}
\end{figure}

Corresponding SHWFS measurements of $\bar{w}_{x}$ are shown in Fig.~\ref{result_para}. Again the measurements are qualitatively consistent with both analytical calculations and phase-field simulations. However, the quantitative agreement is not as good as that for $\bar{w}_{y}$ (Fig.~\ref{result_perp}). In particular, the SHWFS measurement seems to underestimate the distribution of negative $\bar{w}_{x}$ behind the crack tip. We are still investigating the cause, but believe that it may result from tilting of the specimen about the $y$ axis during loading. Such a tilt will cause the front surface of the sample to rotate around the $y$ axis, leading to a change in $\Delta x$ but leaving $\Delta y$ unchanged. In the particular case shown in Fig.~\ref{result_para}(b), the sample likely tilted along the positive direction of $y$ (based on the right-hand rule), adding a positive value to $\Delta x$. We note here, and show in the next section, that this also affects the apparent stress intensity factor $K_\text{I}$ extracted using $\bar{w}_{x}$.

\subsection{Extraction of $K_\text{I}$}
\begin{figure}
\centering
\includegraphics[width=\linewidth]{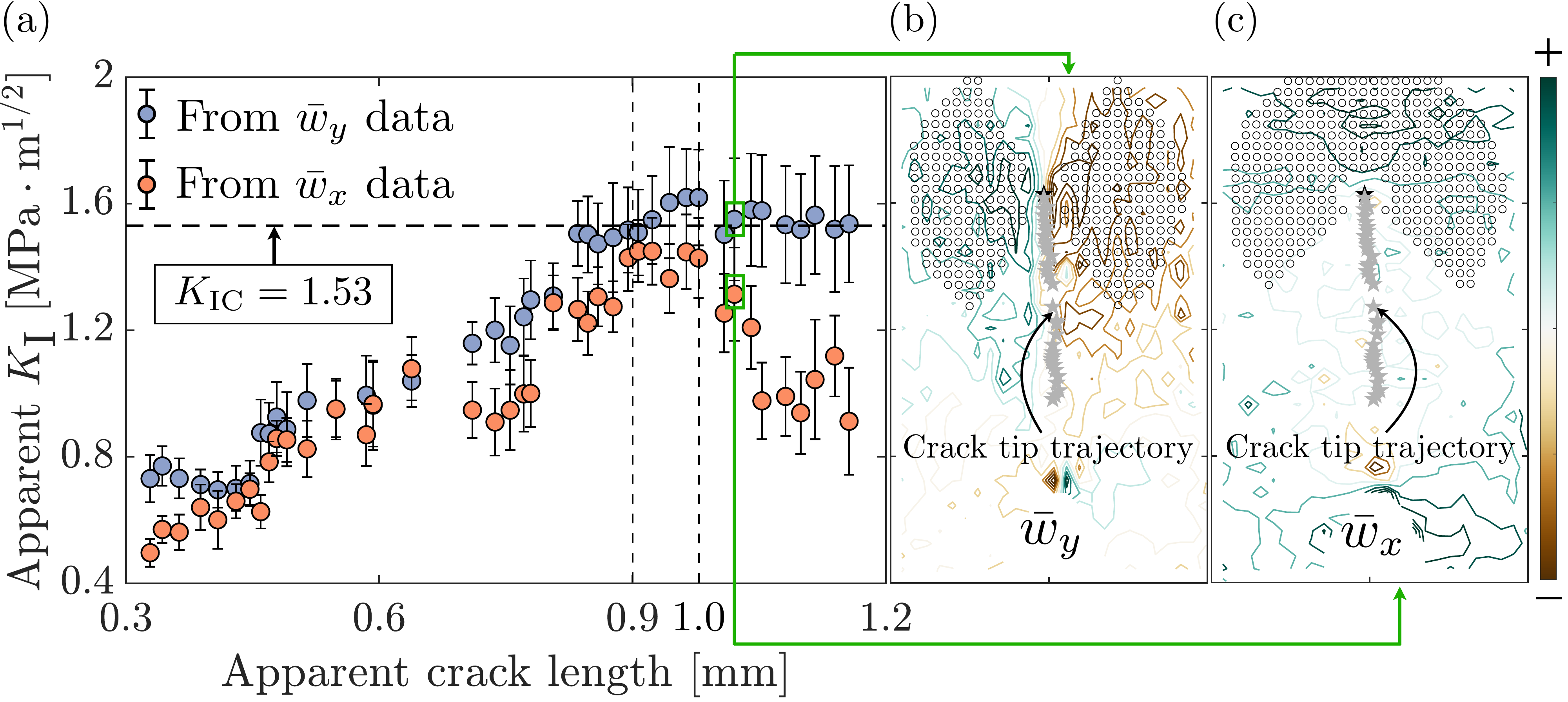}
\caption{(a) Evolution of the apparent stress intensity factor $K_\text{I}$ as a function of the apparent crack length, using both the $\bar{w}_{y}$ data (blue circles) and the $\bar{w}_{x}$ data (red circles). The error bar associated with each data point indicates the scatter coming from varying $\mathbf{P}$ and $r_\text{min}$. (b)~A visualization of a particular snapshot showing the spatial contour plot of $\bar{w}_{y}$, the slope data selected for extracting $K_\text{I}$ (hollow circles), the base location of the crack tip (black star), and the crack trajectory (grey stars). (c)~A similar visualization to (b) but showing the spatial contour plot of $\bar{w}_{x}$.}
\label{k_result}
\end{figure}

Fig.~\ref{k_result}(a) shows how $K_\text{I}$ varies as a function of the apparent crack length. When the crack length is less than \SI{0.9}{mm}, $K_\text{I}$ approaches from the below the critical stress intensity for Macor\textsuperscript{\texttrademark}, $K_\text{IC} = \SI{1.53}{MPa.m^{1/2}}$. This is possibly due to unloading~\cite{dondeti2022cascading} and the time required after crack initiation to establish the singular stress field~\cite{ravi1984experimental}. After the crack length reaches \SI{0.9}{mm}, $K_\text{I}$ is essentially constant. We note that the value of $K_\text{IC} = \SI{1.53}{MPa.m^{1/2}}$ for Macor\textsuperscript{\texttrademark} is that for a quasi-static crack growth, which is a reasonable approximation for these conditions since the crack tip velocity is a small fraction of the Rayleigh wave speed. The uncertainty in $K_\text{I}$ (shown as error bars) in Fig.~\ref{k_result}(a) is computed as the root mean square of the scatter stemming from varying $\mathbf{P}$ and $r_\text{min}$, whose uncertainties are quantified by $r_c$ and $r_d$ as discussed in Section. \ref{extraction_prodecure}.

Although the $K_\text{I}$ trends determined using $\bar{w}_{y}$ and $\bar{w}_{x}$ are similar for crack lengths smaller than about \SI{1}{mm}, they diverge beyond that point. In particular, the value determined from $\bar{w}_{y}$ plateaus at the static critical stress intensity as noted above, but the value determined from $\bar{w}_{x}$ drops. We are still investigating the cause of this drop, but we believe that it may result from the sample tilting around the $y$ axis during loading (see Fig.~\ref{shwfs_sample}(a)). This tilt may be due to a combination of imperfect sample machining (along the sample thickness direction) and sample disintegration due to crack propagation. It only adds changes to $\Delta x$ but not $\Delta _y$, and this can help explain the lack of negative $\bar{w}_{x}$ values measured behind the crack tip, as shown in Fig. \ref{k_result}(c), and in comparison with Fig. \ref{k_result}(b). 

\section{Summary and outlook}\label{summary}
In this work, we demonstrate a new approach for measuring stress intensity during dynamic fracture of brittle solids using a Shack-Hartmann wavefront sensor (SHWFS). With the help of a high-speed camera and high-intensity pulsed laser illumination, we have successfully detected the surface slope ($\bar{w}_x$ and $\bar{w}_y$) evolution during the dynamic fracture of a millimeter-scale Macor\textsuperscript{\texttrademark} sample with a spatial resolution of $\SI{45}{\micro m}$ and a temporal resolution of $\SI{0.2}{\micro s}$. The value of the stress intensity factor ($K_\text{I}$) we extract from SHWFS measurement of $\bar{w}_y$ also agrees with the static fracture toughness of Macor\textsuperscript{\texttrademark}, provided the crack propagates over a distance larger than about \SI{1}{mm} after initiation. Compared to a typical DGS apparatus, the SHWFS apparatus has improved spatial resolution while requiring less physical space to implement. As such, SHWFS can be especially useful for small-scale fracture experiments and is ideal for multi-modal studies, particularly in combination with x-ray phase contrast imaging (XPCI) performed at synchrotron facilities. One potential application in this regard is studying the dynamic fracture of materials under high-rate loadings using miniaturized desktop Kolsky bar setups \cite{akhondzadeh2023direct} at synchrotron facilities \cite{leong2018quantitative}. In addition, unlike DIC or DGS, SHWFS does not require the preparation of speckle patterns (which is particularly challenging for small-scale experiments), and it can be used on any material so long as the surface can be polished to a mirror-like finish. The setup of SHWFS can also be modified to study transparent materials without the need for coating a specimen's sample to be reflective, in a way similar to transmission-mode DGS \cite{periasamy2012full}. As such, the flexibility of SHWFS with regard to sample preparation also makes it adaptable to studying a broader range of brittle materials. 
We discuss two directions for improvement:
\begin{itemize}
\item Increasing the resolution of the measurement. There are two aspects in this direction: increasing the smallest detectable deformation (i.e., sensitivity) and increasing the spatial resolution of detection. The smallest detectable deformation is related to the shift of the reflected spots on the camera sensor. Given a fixed field of view determined by the microscope objective, the shift is tied to the focal length (denoted as $f$) of the multi-lens array (MLA), with a larger focal length producing a larger shift (Eqn.~\ref{wx} and \ref{wy}).  The spatial resolution is tied to the pitch of the MLA (denoted as $a$, essentially how large each microlens is). Since each microlens performs an average measurement of surface slope across its area, we will want an MLA with a small pitch (i.e., a smaller microlens) to increase the spatial resolution. However, diffraction effects limit the smallest pitch (denoted as $a$) and the largest focal length (denoted as $f$) of an MLA. Therefore, a tradeoff exists between a small $a$ and a large $f$. Specifically, the Fresnel number $\frac{a^2}{f \lambda}$ must be larger than unity where $\lambda$ is the wavelength. Our current setup ($a = \SI{150}{\micro m}$, $\lambda = 640$\,\,nm, and $L = 5.6$\,\,mm) leads to a Fresnel number of about 6.2. Choosing a different MLA for our setup may lead to a better detection sensitivity with minimal sacrifice to the spatial resolution or vice versa. Nevertheless, it is important to note that the choice of MLA is application-specific: depending on the specific experiment details (such as material type and size), a different MLA may be needed for the desired detection sensitivity and spatial resolution. Lastly, note that we also have the option of enlarging the field of view (by using a microscope objective with a smaller magnification, see Eqn. \ref{conversion}), but this can require a larger sample and lead to a change of the loading apparatus.

\item Increasing the accuracy of the crack tip position determination. We have shown that an accurate estimation of $K_\text{I}$ depends heavily on accurately identifying the crack tip location (as also highlighted in \cite{taylor2023image, gupta2023identifying}). This is quite challenging to do at the millimeter scale based solely on the distribution pattern of the surface slope, especially given the presence of noise in experimental measurements. In this regard, one possibility is exploiting the distribution property of the out-of-plane asymptotic field, in a way the extends the novel approach proposed by \cite{gupta2023identifying} which takes advantage of the separability of the in-plane asymptotic field (e.g., the in-plane displacement field measured by DIC). Another possibility is integrating SHWFS and x-ray phase contrast imaging (XPCI), with the latter providing more accurate time-resolved identification of the crack tip location. In XPCI, phase perturbation introduced by the sample is exploited to modulate the intensity recorded at the image detector plane~\cite{endrizzi2018x}. Since the surface of a crack induces a steep phase gradient, a phase-contrast image can have a significantly enhanced contrast compared to conventional radiography. XPCI has already been used to characterize the crack dynamics within different brittle materials from ceramics~\cite{leong2018quantitative} to glass \cite{parab2014observation, kang2020crack} taking advantage of spatially-coherent high-energy X-rays~provided by synchrotron facilities. 
\end{itemize}

\section{CRediT Authorship Contribution Statement}
\textbf{L Li}. Conceptualization, Methodology, Experiment, Numerical Simulation, Formal Analysis, Investigation, Visualization, Writing -- Original \& Draft. \textbf{V. Kilic}. Methodology, Experiment, Formal Analysis, Investigation, Visualization, Writing -- Original \& Draft. \textbf{M. Alemohammad}. Methodology, Experiment, Writing --  Review \& Editing. \textbf{KT Ramesh}. Conceptualization, Resources, Supervision, Funding Acquisition, Project Administration, Writing -- Review \& Editing. \textbf{MA Foster}. Conceptualization, Methodology, Resources, Supervision, Project Administration, Writing -- Review \& Editing. \textbf{TC Hufnagel}. Conceptualization, Resources, Supervision, Funding acquisition, Project Administration, Writing -- Review \& Editing.

\section{Acknowledgements}
We thank Dr. Jason Harris, Dr. Charlene Smith, and Dr. Xinyi Xu from Corning Inc. for stimulating discussions on glass ceramics, as well as Nand Vommi for his help with sample preparation, Justin Moreno and Matthew Shaeffer for their assistance to our experiments, and Dr. Chengyun Miao for insightful discussions on DGS. We gratefully acknowledge financial support provided by the Corning Research and Development Corporation. The project was sponsored in part by the Department of Defense, Defense Threat Reduction Agency under the Materials Science in Extreme Environments University Research Alliance, HDTRA1-20-2-0001. The content of the information does not necessarily reflect the position or the policy of the federal government, and no official endorsement should be inferred. 

\appendix

\section{Measuring the indentation speed}
\begin{figure}[h]
\centering
\includegraphics[width=\linewidth]{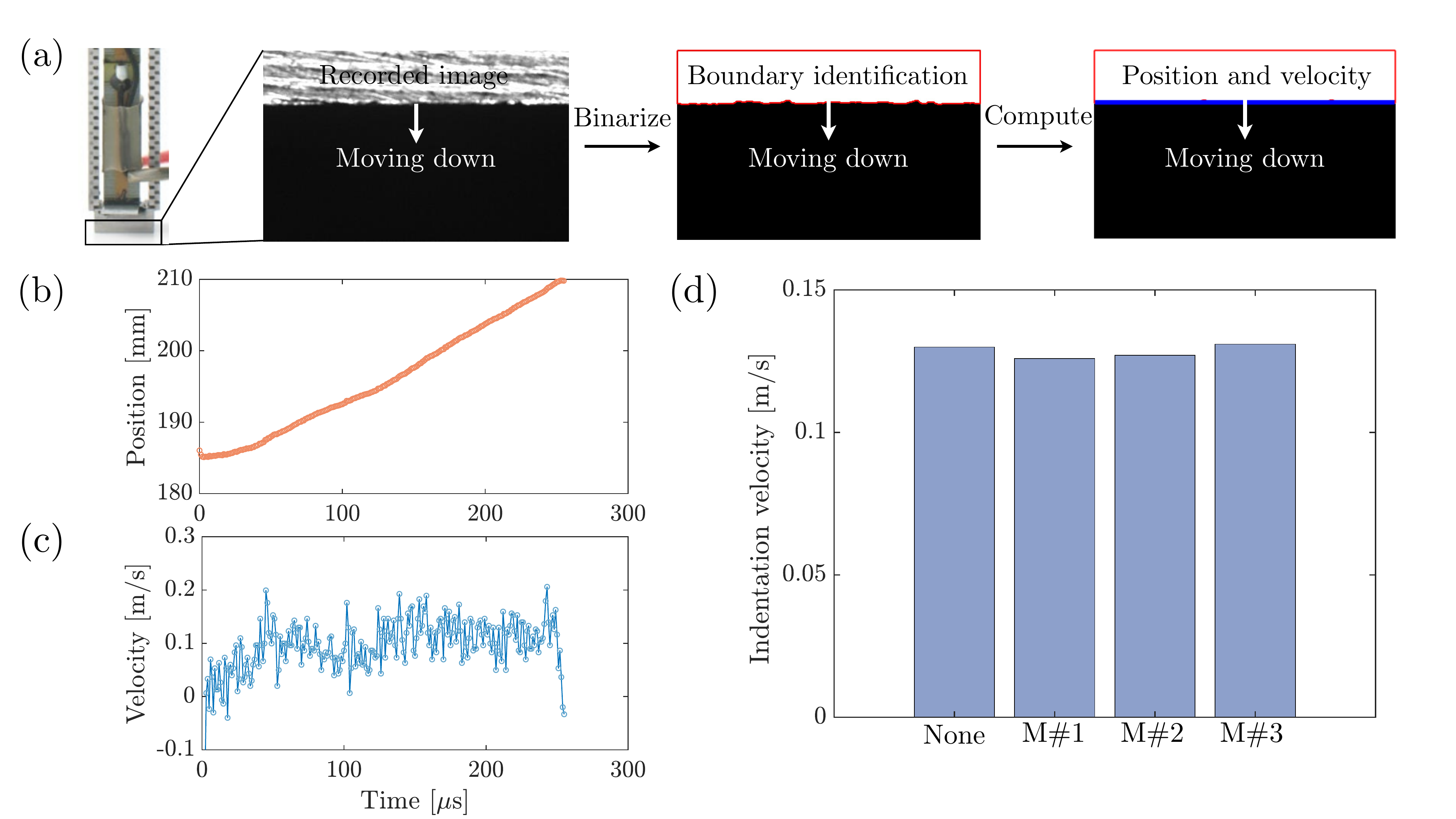}
\caption{(a)A schematic showing the steps of calculating the indentation speed. (b) The position evolution of the actuator's bottom edge as a function of time without indenting a sample. (c) The corresponding speed evolution of the actuator's bottom edge as a function of time. (d) The time-averaged indentation speed without and with indenting a sample (which is repeated three times).}
\label{indetationvelocity}
\end{figure}

We recourse to image analysis to measure the indentation speed of the piezo actuator (which drives the indenter), following the same procedure the lead author has used to measure the angle of repose of granular materials \cite{li2020identifying}. We record the motion of the bottom edge of the piezo actuator using a high-speed camera, shown in Fig. \ref{indetationvelocity}(a), and we compute the position variation, shown in Fig. \ref{indetationvelocity}(b) as a function of time by binarizing each image and finding the boundary of the actuator in each image. From the position data, we can calculate the instantaneous velocity, shown in Fig. \ref{indetationvelocity}(c), using a finite-difference scheme. We see that the velocity is fairly constant. We calculate the average velocity of the actuator following the above procedure, both with and without indenting a sample. Fig. \ref{indetationvelocity}(d) shows the result. In short, the indentation velocity stays largely unchanged, equaling around 0.13 m/s regardless of whether indenting a sample or not.

\section{Variational phase-field approach to fracture}
\begin{figure}[h]
\centering
\includegraphics[width=\linewidth]{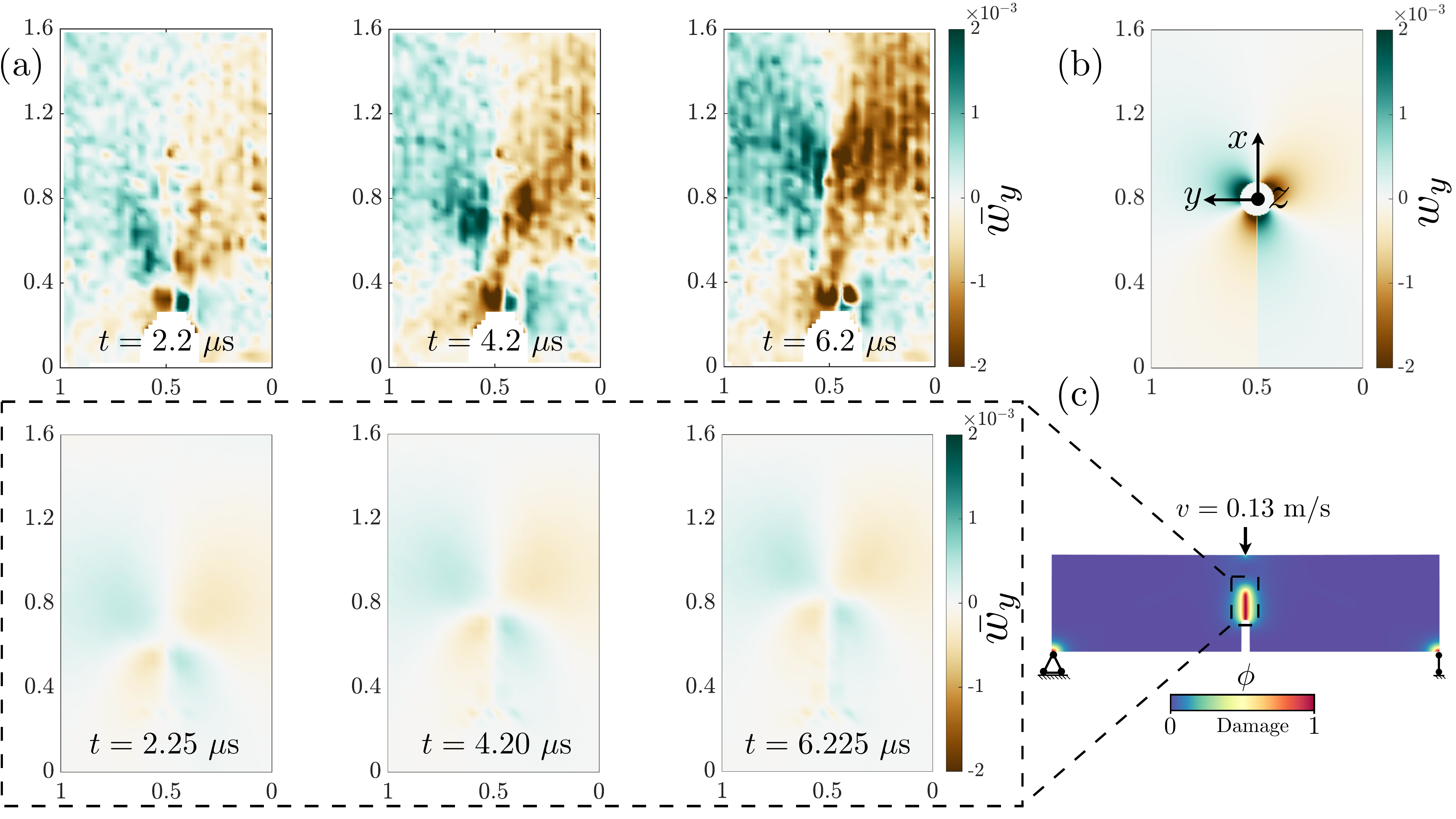}
\caption{A similar figure to Fig. \ref{result_perp} but with plots shown in (c) obtained from $\ell = 0.15$ mm.}
\label{perp_largeL}
\end{figure}

\begin{figure}[H]
\centering
\includegraphics[width=\linewidth]{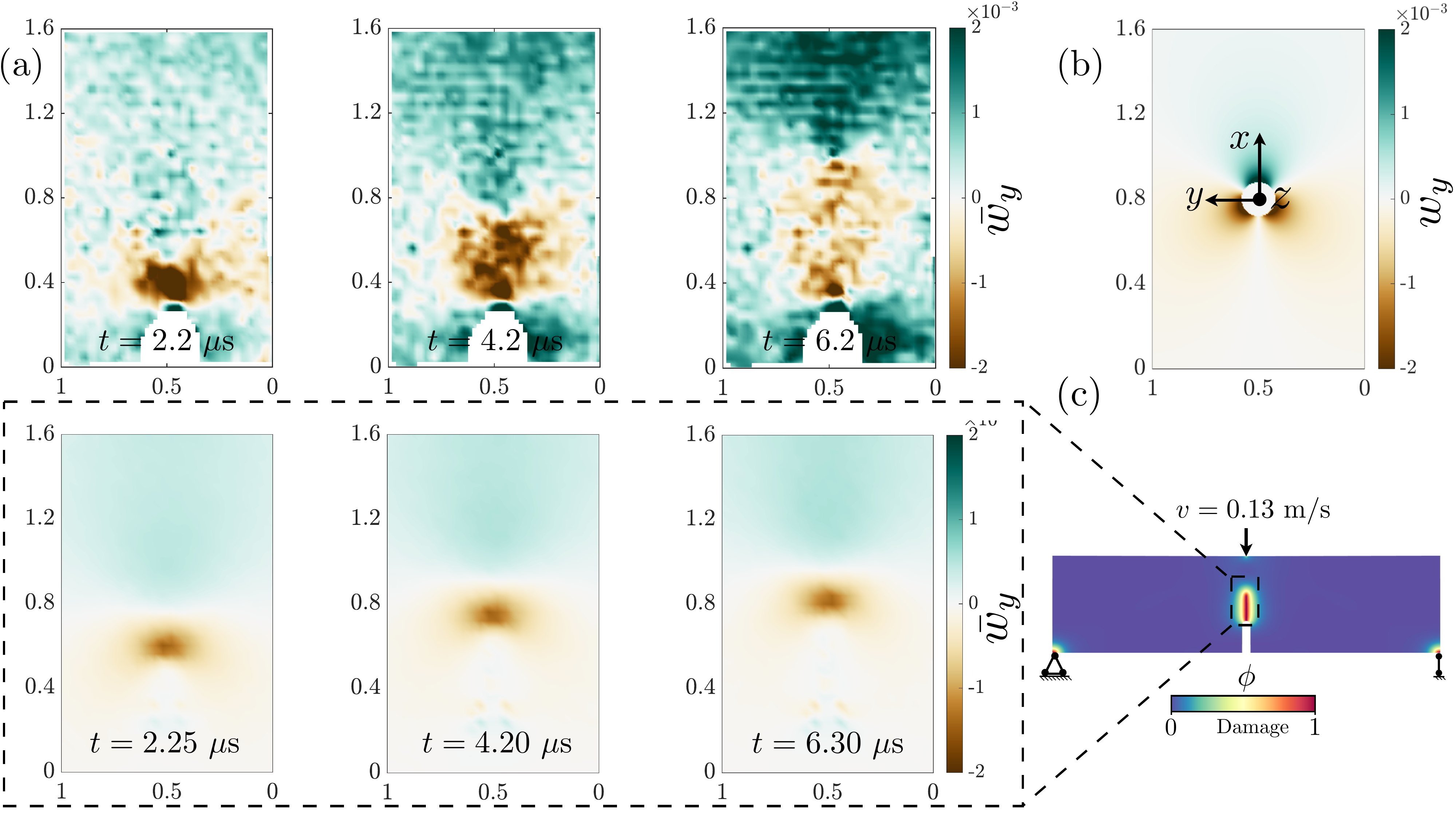}
\caption{A similar figure to Fig. \ref{result_para} but with plots shown in (c) obtained from $\ell = 0.15$ mm.}
\label{para_largeL}
\end{figure}

The variational phase-field method to fracture \cite{bourdin2008variational} uses a scalar field $\phi \in [0, 1]$ to describe the spatial distribution of damage ($\phi = 0$ being intact and $\phi=1$ being completely damaged) of a simulation domain through a length scale parameter $\ell$. Consequently, a crack (a sharp interface) is smeared through a smooth variation of $\phi$ over $\ell$. The parameter $\ell$ represents active mechanisms in the process zone, determining the threshold for crack nucleation \cite{bourdin2008variational}. Essentially, to model a dynamic fracture problem, we want to minimize the following incremental Lagrange energy functional $\mathbf{I}_\ell$ by the principle of least action \cite{borden2012phase}:
\begin{align}
\mathbf{I}_\ell(u,\dot{u},\phi) = \int_{t_1}^{t_2}\left\{  \int_{\Omega} \left[  \frac{\rho}{2}|\dot{u}|^2- \mathcal{W}^e(u,\phi)-G_
\text{C}\gamma_{\ell}(\phi,\nabla \phi) +\rho b\cdot u \right] \text{d$\Omega$}  +\int_{\partial\Omega}   t \cdot u \text{dS}    \right\}\text{dt},
\label{pfintegral}
\end{align}
under the constraint $\dot{\phi} > 0$ to account for the irreversibility of a fracture process. Above, $u$ is the displacement field, with $\dot{u} = \frac{\partial u}{\partial t}$ the velocity field, $\rho$ the material density, $\phi$ the phase field parameter indicating the degree of material damage, $\mathcal{W}^e$ the elastic strain energy density, $G_\text{C}$ the critical energy release rate (or fracture toughness) \cite{griffith1921vi}, $\gamma_\ell$ the (regularized) fracture energy density, $b$ the gravitational constant, and $t$ the surface traction. We follow \cite{miehe2010phase} for modeling brittle solids, decomposing the elastic strain energy shown in Eqn. \ref{pfintegral} to a tensile part ($``+$") and a compressive part ($``-"$), with the phase-field acting only on the former:
\begin{align}
\mathcal{W}^e(\epsilon_{ij},\phi) = [(1-k)(1-\phi)^2+k]\mathcal{W}^{e,+}(\epsilon_{ij})+\mathcal{W}^{e,-}(\epsilon_{ij}),
\label{energydecompose}
\end{align}
where $\epsilon^d$ is the $d$-th eigenvalue of $\epsilon$,  $n^d$ is the corresponding eigenvector, $\langle x \rangle_+$ stands for $(x +|x|)/2$, and  $\langle x \rangle_-$ stands for $(x -|x|)/2$ with $|x|$ being the absolute value of $x$. We can then express $\mathcal{W}^{e,+}(\epsilon_{ij})$ and $\mathcal{W}^{e,-}(\epsilon_{ij})$ as the following:
\begin{align}
\mathcal{W}^{e,+}(\epsilon_{ij}) &= \frac{1}{2}\lambda \langle \epsilon_{kk} \rangle_{+}^2+\mu\epsilon^{+}_{kj}\epsilon^{+}_{jl}\delta_{kl}, \label{energyplus}\\
\mathcal{W}^{e,-}(\epsilon_{ij}) &= \frac{1}{2}\lambda \langle \epsilon_{kk} \rangle_{-}^2+\mu\epsilon^{-}_{kj}\epsilon^{-}_{jl}\delta_{kl}, \label{energyminus}
\end{align}
where $\lambda$ and $\mu$ are the Lam\'e constants that can be determined from the Young's modulus $E$ and the Poisson's ratio $\nu$. The (regularized) fracture energy density $\gamma_\ell$ 
 takes the following form:
\begin{align}
\gamma_\ell = \frac{1}{4c_w\ell}\left(w(\phi)+\ell^2|\nabla \phi|^2 \right), \text{with}\, c_w = \frac{1}{2}\,\,\text{which implies}\,\, w(\phi) = \phi^2.
\label{pfdensity}
\end{align}
 Applying the principle of least action to Eqn. \ref{pfintegral} with $\mathcal{W}^e$ and $\gamma_\ell$ 
 expressed by Eqns. \ref{energydecompose}, \ref{energyplus}, \ref{energyplus}, and \ref{pfdensity}, we arrive at the following two governing equations:
\begin{align}
\sigma_{ij,j}+b_i &= \rho \ddot{u}_i,\\
\left[1+\frac{4c_w\ell(1-k)}{G_\text{C}}\mathcal{W}^{e,+}\right]\phi-\ell^2\phi_{,ii} &= \frac{4c_w\ell(1-k)}{G_\text{C}}\mathcal{W}^{e,+},
\label{governing}
\end{align}
where $\sigma_{ij} = \partial \mathcal{W}^{e}/\partial {\epsilon_{ij}}$. We enforce the irreversible growth condition $\dot{\phi} > 0$ using a strain-history field \cite{miehe2010phase} over the simulation domain:
\begin{align}
\mathcal{H}(x,t) = \max_{s \in [0,t]}\mathcal{W}^{e,+}\left(  \epsilon(x,s)  \right)\,\,\forall\, x \in \Omega.
\label{historyfield}
\end{align}
Replacing $\mathcal{W}^{e,+}$ with $\mathcal{H}(x,t)$ in Eqn. \ref{governing} we then want to solve:
\begin{align}
\sigma_{ij,j}+b_i &= \rho \ddot{u}_i, \label{ufield}\\
\left[1+\frac{4c_w\ell(1-k)}{G_\text{C}}\mathcal{H} \right]\phi-\ell^2\phi_{,ii} &= \frac{4c_w\ell(1-k)}{G_\text{C}}\mathcal{H}, \label{phifield}
\end{align}
together with the following Neumann boundary conditions (plus any existing Dirichlet boundary conditions) :
\begin{align}
\sigma_{ij}n_j &= t_i\,\,\text{on}\,\, \partial \Omega,\\
\phi_{,i}n_i &= 0\,\,\text{on}\,\, \partial \Omega.
\end{align}
We solve Eqns. \ref{ufield} and \ref{phifield} weakly follow a standard finite element discretization and calculation procedure, using the alternating minimization (or staggered) scheme that runs parallel on high-performance computer clusters. We use the mechanical properties $E = 66.9\,\, \text{GPa}, K_\text{IC} =  \SI{1.53}{MPa.m^{1/2}}$, and $\nu = 0.29$ in our PF simulations. The notched beam in the simulation shares the same geometry as the experiment specimen, having a notch of $\SI{250}{\micro m}$ in width and 1 mm in length. Since $\ell$ affects crack nucleation, it will affect values of $w_{x}$ and $w_{y}$. Essentially, a smaller value of $\ell$ suggests more delayed crack nucleation, leading to larger values of $w_{x}$ and $w_{y}$ because of a larger stress build-up before crack nucleation. We find $\ell = 0.035$ mm to be a reasonable choice, and the corresponding results are presented in the main text. We pick an element size of $\delta h = 0.015$ mm near the crack propagation region, which is small enough ($\delta h \leq \ell/2$ \cite{miehe2010phase}) to resolve the crack geometry. However, we emphasize that $\ell$ does not change the spatial pattern of $w_{x}$ and $w_{y}$, which is a direct consequence of the balance of linear momentum. Figs. \ref{perp_largeL}(c) and \ref{para_largeL}(c) shows results obtained from $\ell = 0.15$ mm. Compared to Figs. \ref{result_perp}(c) and \ref{result_para}(c), the spatial pattern of $\bar{w}_{x}$ and $\bar{w}_{y}$ stays the same, but the magnitude is smaller.

\bibliographystyle{unsrt}
\bibliography{References}
\end{document}